\documentstyle[paspconf,psfig,epsf,twoside]{article}

\begin{document}

\title{Theoretical Calculations of Atomic Data for Spectroscopy} 
\author{Manuel A.  Bautista}
\affil{Laboratory for High Energy Astrophysics}
\affil{NASA Goddard Space Flight Center} 
\affil{Greenbelt, MD 20771, USA}

\begin{abstract}

Several different approximations and techniques have been developed 
for the calculation of atomic structure, ionization, and excitation of 
atoms and ions. These techniques have been used to compute large amounts of spectroscopic data of various 
levels of accuracy. This paper presents a review of these 
theoretical methods  to help non-experts  
in atomic physics 
to better understand the qualities and
limitations of various data sources and assess how reliable are
spectral models based on those data. 

\end{abstract}

\keywords{atomic processes -- line formation -- X-rays:spectroscopy} 

\section{Introduction} 

Spectroscopic data for atoms and ions fall into the category of
fundamental scientific research and this by itself may be good 
justification for experimental and theoretical work. The
data are also essential to the analysis of spectra of laboratory 
and astronomical plasmas. These applications have been as important
motivation as, if not greater than, the fundamental physics 
For several decades, theoretical atomic physicists have devoted 
great effort to the development of methods and computation of vast 
amounts of data, particularly that needed for the study of optical
spectra.
On the other hand, the progress of the atomic data sets for X-ray 
spectroscopy has been much slower, due in part to the scarcity of 
high quality X-ray spectra to motivate the work. At present, 
the new generation of observatories (like {\it Chandra} and {\it XMM})
and the incoming facilities with the {\it Constellation-X} program 
are starting to provide high resolution, high signal-to-noise spectra 
for which the need of improved atomic data is not just 
granted but urgent.
But the production of high quality atomic data for X-ray spectroscopy 
is far from simple or routine. The high energy conditions of the 
emitting plasmas demand the study of large numbers of energy levels as
well as inner-shell and Auger processes, 
Thus, current theoretical methods and 
computational tools encounter severe difficulties in producing 
high accuracy data. 

The present review of theoretical atomic physics is aimed at astronomers
and spectroscopists in general who need to understand the limitations of atomic
data. In that sense, the present paper tries to give a general, but 
relatively simple,  overview of the
most commonly used approximations and methods for the production of 
spectroscopic data. The review pays particular attention to the
the calculation of atomic structure and the spectral processes of
photoionization and electron impact excitation. Other processes 
like collisional ionization and proton impact excitation, although 
relevant, are not discussed in much detail due to lack of space. 
For discussions on electron-ion recombination see papers by Nahar and 
Savin in this volume.
Also, the present review emphasizes inner-shell processes 
that are most relevant to X-ray spectroscopy.

From the point of view of the kinds of atomic data needed for the
analysis of spectra there are two general atomic physics problems to be
solved:

\begin{tabular}{ll}
1) Atomic structure & - Energy Levels \\  
                    & - Radiative transition rates \\ 
                    & - Autoionization rates \\ 
 & \\  
2) Scattering       & - ion + photon \\ 
                    & - ion + charged particle 
\end{tabular}

The atomic structure problem is concerned with the computation of 
energy levels and rates 
of spontaneous transition rates among levels of the same ion 
(transition rates or $A$-values), and from autoionizing levels of 
one ion to levels of the next ionized species (autoionization
rates). The scattering problem, relevant to the formation of spectra of 
warm and hot plasmas, has to do with all different processes that 
can occur after a collision of an ion with a photon or with a charged 
particle.
The most important physical processes that occur after 
ion scattering are illustrated in Figure 1.
 
\begin{figure}[Hbt]
\hskip 1in\psfig{file=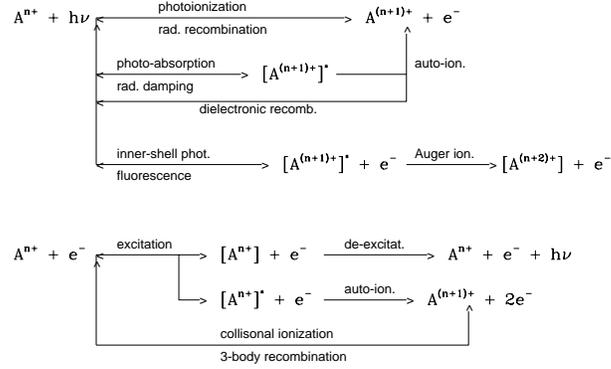,width=4in,angle=-90 }
\caption{Physical processes after photon-ion and electron-ion
scattering. The brackets indicate excited states of the ion and 
brackets and asterisc indicate autoionizing states.}
\end{figure}

\section{Atomic Structure Calculations}

From the point of view of quantum mechanics 
the physical representation of ions and spectral processes comes down to the 
fundamental problem of solving the Schr\"odinger equation 
\begin{equation}
H\Psi_i = E_i \Psi_i.
\end{equation} 
Here, $i$ represents the set of quantum numbers necessary to describe 
the system, $\Psi_i$ are the wavefunctions of the system, $E_i$ are the 
eigenvalues, and $H$ is the Hamiltonian which, for a
non-relativistic system,  can be 
written as
\begin{equation}
H=\sum_{i=1}^N {p_i\over 2m_e} - \sum_{i=1}^N {Z e^2\over r_i}
+ \sum_{i\neq j} {e^2\over |r_i-r_j|},
\end{equation}
where the first term is the sum of the kinetic energies of all
electrons, the second term is the potential energy due to Coulomb 
attraction of all electrons by the nucleus, the third term is the energy
due to electrostatic repulsion between the electrons, and $N$ is the total number of electrons in the system.     
The previous equation neglects the spin of the electron, which can be 
considered separately in 
non-relativistic systems.

The presence of the two-electron operators $1/|r_i-r_j|$ makes it 
impossible to obtain exact solutions to the Schr\"odinger equation 
for the $N$-electron system. On the other hand, the two-electron 
operator should never be neglected as it is often comparable in
magnitude to the Coulomb attraction term or one-electron operator. 
One can show
\begin{equation}
{two-electron\  term\over one-electron\  term} \equiv \xi \simeq  
{1\over 2}{N(N-1)\over ZN}.
\end{equation}
For a neutral ion ($Z=N$), $1/4\le \xi \le 1/2$.

Approximate methods for solving 
Equation (1) replace the two-electron terms by 
approximate one-electron potentials to give an effective Hamiltonian 
of the form 
\begin{equation}
H^{eff}=\sum_i^N H_i^{eff} = -\sum_{i=1}^N\left[{1\over 2}{p_i^2\over m_e}
+{Ze^2\over r_i}-V_i^{eff}(r_i)\right].
\end{equation}
The exact eigenfunctions of $H^{eff}$ can now be constructed from 
products of single-electron eigenfunctions of $H_i^{eff}$. Since 
$H$ commutes with angular momentum and spin operators ($L^2, L_z,
S^2, S_z$) the eigenfunctions of $H^{eff}$ are required to be
eigenfunctions of these operators as well.

For highly ionized ions ($Z \gg N$), the inter-electronic repulsions 
are only small perturbations relative to the much stronger nuclear 
central potential. Then, 
 it is a good approximation to
choose $V_i^{eff}$ as a {\it central field potential}.
If this can be achieved the orbital wavefunction may be written
explicitly as products of spherical harmonics, spin functions, and 
radial functions. The advantage of using central field orbitals   
cannot be overestimated. The effect is to reduce the computational
problem from solving $N$ coupled integro-partial-differential equations
(one for each orbital) in four variables to $N$ coupled
integro-differential equations in a single radial variable. 

In general $V^{eff}$ does not need to be of central field type,
but the advantages of a central potential are so great that essentially
all 
methods for atomic structure calculations use spherically
averaged potentials even for neutral and open-shell systems.
There are several techniques regularly used in the atomic structure calculations.
The most important are: use of model potentials, methods based on the Hartree-Fock
theory, semiempirical methods, perturbation methods, and the R-matrix
method in the close coupling formalism. These methods,
with the exception of the R-matrix,
are discussed below followed by a discussion of two particularly important
physical effects, i.e. configuration interaction and relativistic
effects. The R-matrix method will be described in Section 3.5.

\subsection{Model potentials in atomic structure}

Alkali atoms and ions are good systems for testing model potentials
since, by having  a single valence electron, their analysis leads to a
single equation for the one-electron wavefunction. For these ions one 
would expect the potential for large $r$ to approach
\begin{equation}
V^{eff}(r)=-{(Z-N_c)\over r},
\end{equation}
where $N_c$ is the number of core electrons. The solution 
of the Schr\"odinger equation based on this potential is discussed by 
Bates and Damgaard (1949). In spite of the simplicity of this 
potential it gives accurate oscillator strengths in many cases, but 
it is not always reliable (Bromander \etal 1978).

A large variety of model potentials have been constructed from the 
assumptions that they behave as shown in Equation (5) for $r> R$,
for some suitable $R$, while for $r$ approaching 0 the potential could
be infinite, equal to a finite boundary value, or even zero.
For a review of model potentials in atomic structure see Hibbert (1982).

A generally applicable potential is the modified Thomas-Fermi 
potential by Eissner and Nussbaumer (1969). This potential is implemented 
in the widely used computer program SUPERSTRUCTURE (Eissner, Jones, and
Nussbaumer 1974). This program uses a Thomas-Fermi-Dirac type of central
potential to generate one-electron orbitals. 

This potential differs from that in Equation (5) not only in the 
added sophistication of the model, but also in that this one depends 
on the angular momentum of the valence electrons.
SUPERSTRUCTURE is quite efficient and can
provide relatively accurate results ($\sim 1\%$ for energy levels and
$\sim 10\%$ for oscillator strengths). Recently the code has been 
extended to calculate autoionization and dielectronic recombiantion
data (AUTOLSJ by Bely-Dubau 1982 and AUTOSTRUCTURE by Badnell 1985,
unpublished) has 
extended the code to calculate autoionization rates and 
dielectronic recombination. 

\subsection{Atomic structure calculations base on the Hartree-Fock 
Formalism} 

In contrast to the local central-field potentials, the Hartree-Fock (HF)
method entails the computation of the one-electron orbitals in the
non-local potential (direct and exchange) generated from electronic
orbitals in a self-consistent manner using the variational principle.
More extensive discussion of the HF
method can be found in Hartree (1957), Slater (1960), Froese Fischer
(1977), Cohen and McEachran (1980).

Early HF calculations indicated  that in going from the ground state 
of beryllium $1s^22s^2\ ^1S$ to the first excited states $1s^22s2p\
^{1.3}P^o$, the $1s$ orbital remains almost unchanged. This led 
Fock (1933) to realize that accurate wavefunctions could be found 
by varying only valence orbitals while keeping the orbitals for the 
core fixed. This approach is known as the {\it frozen core} (FC)
approximation. 
  
In spite of the success of the FC approximation, this is not always 
directly applicable, such as in the case of configurations containing
equivalent electrons. The simplest example of such configurations 
is the ground state of helium $1s^2\ ^1S$, which has two equivalent 
electrons, both described by a single radial function $P_{1s}(r)$. Yet,
if one identifies one of the electron as ``core" and the other as 
``valence" they require different radial functions. Evidently, these 
two $1s$-orbitals cannot be made orthogonal. 
The use of nonorthogonal orbitals to solve the atomic structure problem
leads to equations much more complex than otherwise. 
Calculations with such nonorthogonal orbitals have been carried out only for
a few systems (e.g. Pratt 1956; Froese 1966; Jucys 1967). 

The FC approximation yields goods results for simple systems, but 
for the mayority of cases it is necessary to include electronic 
configuration interactions like in the multiconfiguration 
Hartree-Fock (MCHF) or in the superposition of configurations 
(SOC) methods. 
For example, for boron-like cores a possible
two-configuration core wave function is 
\begin{equation}
\Psi(^1S)=c_1\Psi_1(1s^22s^2\ ^1S) + c_2 \Psi_2(1s^22p^2\ ^1S).
\end{equation} 
Clementi and Veillard (1965) showed that including $\Phi(1s^22p^2\ ^1S)$
accounts for most of the correlation energy error for all values of $Z$.

The MCHF (e.g. Froese Fischer 1977) method computes orthonormal orbitals
self-consistently in an iterative fashion for every choice of the $\{c_i\}$
coefficients. Then, these coefficients are varied and 
the radial wavefunctions are recomputed until sufficient degree of 
convergence is achieved.
This makes the method potentially very accurate, but computationally
lengthy.

Based on the MCHF technique, Hibbert (1975) developed 
the SOC method in which analytic 
radial functions depending on variational parameters are used. 
Core wave functions of the kind shown in Equation (7) 
are used and the $\{c_i\}$ coefficients are determined for a given 
choice of the radial wavefunction parameters. 
Then, the radial wavefunctions are changed and the coefficients  
determined again until the description of the atom converges
satisfactorily.
The SOC method is somewhat more efficient than the MCHF technique and 
is able to provide similarly accurate results.
The main implementation of the SOC method is in the computer program 
CIV3 (Hibbert 1975).

Another approach used to improve calculations of atomic structure
arises from the realization that a single valence electron may polarize
the spherically symmetric core, including a dipole moment at the 
nucleus of the form $\alpha/r^2$, where $\alpha$ is the dipole
polarizability of the core (Hartree 1957). This gives rise to an 
additional attractive field with a long-range $r$ dependence which is experienced by the valence
electron, and at large distances may be represented by a ``polarization
potential"
\begin{equation}
V_{pol}=-\alpha/2r^4.
\end{equation}
This potential is singular at $r=0$, but it can be modified to 
avoid this singularity as (Biermann and Trefftz 1953)
\begin{equation}
V_{pol}=-(\alpha/2r^4)[1-exp(-x^p)],\   \  x=r/\rho\ .
\end{equation}
The use of a $V_{pol}$ in the orbital equations for valence electrons
can be formally justified (Caves and Dalgarno 1972), and it has 
been shown (e.g. Cohen and McEachran 1980, and references therein)
to provide a very satisfactory description of simple spectra.
 However, the method has the disadvantage that the choice of the parameters $p$ and
$\rho$ in Equation (9) remains completely {\it ad hoc}.

\subsection{Semiempirical methods}

Semiempirical methods try to compute the atomic structure of ions 
by solving simplified forms of the HF equations. One of the first
examples of this approach is the Hartree-Fock-Slater (HFS) scheme, which 
consists in replacing the two-electron non-local exchange terms of the Hamiltonian
by the statistical potential function
\begin{equation}
V_{xs}(r)=-{3\over 2}\left({24\over \pi}\rho\right)^{1/3},
\end{equation}
where $\rho$ is the local electron density $\rho(r)$ of the atom.

Later, Cowan (1967) modified the HFS approach to write the total 
potential of the ion as 
\begin{equation}
V^i(r)=-2Z/r+V_e(r)+V_{xs}(r),
\end{equation}
where $V_e$ is the potential energy for the density $\rho-\rho_i$ of
electrons other than $i$ and $V_{xs}$ 
is an exchange term of the form
\begin{equation}
V_{xs}=-k_1f(r)\left[{\rho'\over \rho'+k_2/(n-l)}\right]\left({\rho'\over \rho}\right)\left({24\over
\pi}\rho\right)^{1/3},  
\end{equation}
where $k_1$ and $k_2$ are constants, $\rho'$ is the electron density less the
densities of the electron $i$ and of the electron with which it is
paired, and 
\begin{equation}
f(r)= \left\{ \begin{array}{ll}
        1&   r\ge r_0 \\
         1+k(1-r/r_0) &  r< r_0
     \end{array} \right.
\end{equation}
for suitable $k$ and $r_0$.  

This definition for the atomic potential is mostly 
empirically motivated and requires preconceived wave functions
with which to construct the electron density functions $\rho$ and
$\rho'$. Once these functions are created the atomic structure 
equations are solved in an iterative fashion. Then, the accuracy 
of the results is assessed upon the agreement between obtained energy levels and
oscillator strengths with a  given sample of experimental 
values. The method has the advantage of being quite efficient,
but it requires a lot of care in 
the construction of the initial electron density distribution.
Further, it is difficult to estimate the accuracy of any given 
calculation except by the observed agreement of a limited sample 
of data with experimental values. 

The Cowan code has been widely used by many researchers like Kurucz
(e.g. Kurucz 1988; Kurucz and Peytremann 1975),
who computed millions of energy levels and oscillators strengths 
for most ions of astrophysical interest.

\subsection{Perturbation treatments}

Perhaps the best known perturbation treatment is the Z-expansion
method which was first introduced by Hylleraas (1930). The main 
idea behind this method is that one can rewrite the $N$-electron 
atom Hamiltonian of Equation (2) in terms of units of length of
$Z$ atomic units (au) and unit of energy of $Z^2$ au (thus different 
units for each $N$-electron ion) to obtain 
\begin{equation}
H=\left[{1\over 2} \sum_i^N {p^2\over m_e}-\sum_i^N {1\over r}\right] 
+ Z^{-1}\left[\sum_{i\neq i} {1\over |r_i-r_j|}\right].
\end{equation}
Then, for highly ionized systems ($Z\gg N$) one can divide this   
Hamiltonian as indicated by the brackets into a one-electron zero 
order Hamiltonian ($H_0$) and a two-electron perturbation ($H_1$) 
with the expansion parameter $Z^{-1}$. Perturbation theory leads 
to the expansion for the energy and wavefunction of the form
\begin{equation}
E=E_0Z^2+E_1Z+E_2+E_3Z^{-1}+...
\end{equation}
\begin{equation}
\Psi=\Psi_0Z^{3/2}+\Psi_1Z^{1/2}+\Psi_2Z^{-1/2}+...
\end{equation}
Here, $H_0$ is the sum of hydrogenic Hamiltonians, so $\Psi_0$ is 
a linear combination of Slater determinants of hydrogenic orbitals. 
Subsequent orders of the Hamiltonian and wavefunctions are found 
as in standard perturbation theory.

The Z-expansion method is conceptually simple, but computationally
lengthy when trying to go beyond the first perturbation term. One
advantage of the method is that a single calculation provides results
for a whole isoelectronic sequence, but accuracy is 
normally restricted to highly ionized ions. Another problem occurs 
in cases of strong configuration interaction (see Section 2.5) 
where states are labeled by the single configuration with the largest 
mixing coefficient, thus zero order mixing does not represent the 
physical state being considered. 

Several extensions of the Z-expansion method have been developed 
(see Cohen 1988, Crossley 1969, and references therein) and a lot of data for
astrophysical applications has been published in recent years by
Vainshtein, Safronova, and collaborators (e.g. Vainshtein and Safronova
1980; Safronova \etal 1998).  

\subsection{Additional considerations}   

\subsubsection{Configuration interaction}

Configuration interaction (CI) can be seen as a way to correct for 
Hartree's single-electron orbital approximation that
pictures each electron moving individually in the field of the 
nucleus screened by the other electrons. Each electron is described by a
single wavefunction and the whole atom is described by the Slater
determinant of these wavefunction (e.g. Slater 1960). This model admits
the familiar configuration description of, for example, the ground state
of Be $1s^22s^2(^1S)$. This kind of description is often not very
accurate. Much better wavefunctions may be obtained from
linear combinations of single configuration 
wavefunctions of the same total angular momentum and spin symmetry
(Condon and Shortley 1935).
This approach is named CI. 

Going back to the case of the ground state of Be, a good CI description
may be        
\begin{equation}
\Psi(^1S)=\alpha \Psi(1s^22s^2\ ^1S)+\beta \Psi(1s^22p^2\ ^1S),
\end{equation} 
where $\alpha$ and $\beta$ are the so-called mixing coefficients.

Such CI representation has important effects on atomic quantities like 
oscillator strengths. For example, in the transition probability of the
resonant transition $1s^22s^2(^1S)-1s^22s2p(^1P^o)$. In the single
configuration model the transition probability is determined by the
matrix element $<2s|r|2p>$. On the other hand, in the two-configuration
description of the ground $^1S$ state, the matrix element becomes
$(\alpha \pm \beta)<2s|r|2p>$. Depending on the relative signs of 
the mixing coefficients and the $\pm$ sign, the transition may be
strengthened or weakened, but which way the value will go is generally
impossible to know without a full solution of the atomic structure
problem.

CI is very important in the representation of the majority of atomic
systems and is a standard capability of codes like SUPERSTRUCTURE, 
CIV3, Cowan's code, and Froese Fischer's MCHF code.

\subsubsection{Relativistic Effects}

So far we have neglected relativistic effects in atomic structure
calculations and we have considered oscillator strengths for dipole 
allowed transitions under LS coupling. In order to consider forbidden transitions it is
necessary to use $jj$ (or at least intermediate) coupling, and this
requires some treatment of relativistic effects. For heavy atoms and 
ions relativistic effects are important even for allowed transitions.
There are two ways of treating relativistic effects: by the addition of
Breit-Pauli operators to non-relativistic equations, or by the fully
relativistic Dirac formalism. The Breit-Pauli operators are seven and
each one accounts for a specific physical effect (Bethe and Salpeter
1957). These operators are:
(1) the mass operator which gives the correction due to the 
relativistic variation of mass with velocity; (2) the Darwin term
which is characteristic of the Dirac theory but has no obvious physical 
interpretation; (3) the spin-orbit coupling term which arises from the
interaction of spin and orbital magnetic moments of each electron; (4)
the spin-other-orbit term which is due to the interaction between 
the spin of one electron with the orbital magnetic moment of another
electron; (5) and (6) are spin-spin coupling terms that describe the
interactions between the spin magnetic moments of pairs of electrons;
(7) the orbit-orbit coupling term which accounts for the interactions 
between the orbital magnetic moments of pairs of electrons.

The expectation value of the orbit-orbit coupling term cannot be calculated accurately in the 
one-electron approximation. Moreover, experience suggests that this term is
rather small for all ions of interest to the present discussion (Ufford
and Callen 1958; Cohen and McEachran 1980). 
The mass and Darwin terms contribute 
to the overall energy shift, but do not break LS coupling. 
The spin-spin, orbit-orbit, and spin-other-orbit operators serve to
split energy terms into fine structure levels.

Most current codes for atomic structure calculations such as
SUPERSTRUCTURE and CIV3 use the Breit-Pauli
approximation to account for relativistic effects. A more precise
treatment of relativistic effects using the Dirac formalism has been
implemented in the computer package GRASP (Dyall \etal 1989).

\section{Scattering Calculations}

In considering the problems of photoionization and excitation or ionization
by electron impact it is customary to think of the residual ion
as the ``target" or ``core" with $N$ electrons and the incoming/outgoing
(electron impact/photoionization) as ``free electron". Thus, the 
wavefunctions of the $(N+1)$-electron system can be expanded in terms
of products of wavefunctions of the core ($\phi_i$) and those of the 
electron ($\theta_i$), i.e.
\begin{equation}
\Psi=\sum_i \phi_i(x_1,...,x_N)\theta_i(x_{N+1}).
\end{equation} 
Substitution of (18) in Equation (2) yields the system of coupled
equations
\begin{equation}
-(\nabla^2+k_i^2)\theta_i(x)+\sum_{i'}V_{ii'}\theta_i(x)=0
\end{equation}
where
\begin{equation}
V_{ii'}(x)=-{2Z\over r_1}\delta_{ii'}+\int \phi^*(x_1,...,x_N)
  \times \sum_{n=1}^N {2\over
r_{N+1,n}}\phi(x_1,...,x_N)dx_1...dx_N
\end{equation} 
and $k_i^2$ is defined by  
\begin{equation}
E=E_i(N)+k_i^2.
\end{equation}
It is important to notice that these equations have to be solved 
for every value of the energy ($k^2$) and of the total angular momentum ($L$) of the 
$(N+1)$-electron system.
This last condition motivates the so-called {\it partial waves expansion}
in which all states of definite angular momentum of the free      
electron are considered separately. Furthermore, the calculated collision 
strength is 
divided into partial waves as
\begin{equation}
\Omega=\sum_{l=0}^\infty \Omega_l
\end{equation}

There are several methods to solve 
the scattering problem and calculate the cross sections necessary 
for practical applications. Below is a discussion of the most widely 
used methods ordered according to their level of sophistication, starting
with the simplest method.
This discussion is followed by a study of several physical effects that
affect the cross sections. 
The techniques are: the central field approximation, 
used in photoionization calculations; the  Gaunt factor and the Coulomb-Born approximation, 
used for electron impact excitation;
 and the Distorted Wave and the R-matrix methods, used for both photoionization and electron 
impact excitation and ionization, as well as atomic structure.

\subsection{The central field approximation}

This is the simplest approximation used to solve the multi-electron
problem. As discussed above, in the context of atomic structure 
calculations, it is assumed that inter-electronic couplings are small and
can be treated as small perturbations with respect to an effective 
central field potential that includes the attraction by the nucleus 
less the screening by the core electrons. A further substantial
simplification of the problem results by treating {\it subshells}
of equivalent electrons (having the same values of $n$ and $l$, but 
different values of $m$ and $s$) by means of a single radial function.
This has the effect of reducing the number of coupled radial equations
to only one for each subshell.

\begin{figure}[Htb]
\hskip 1in\psfig{file=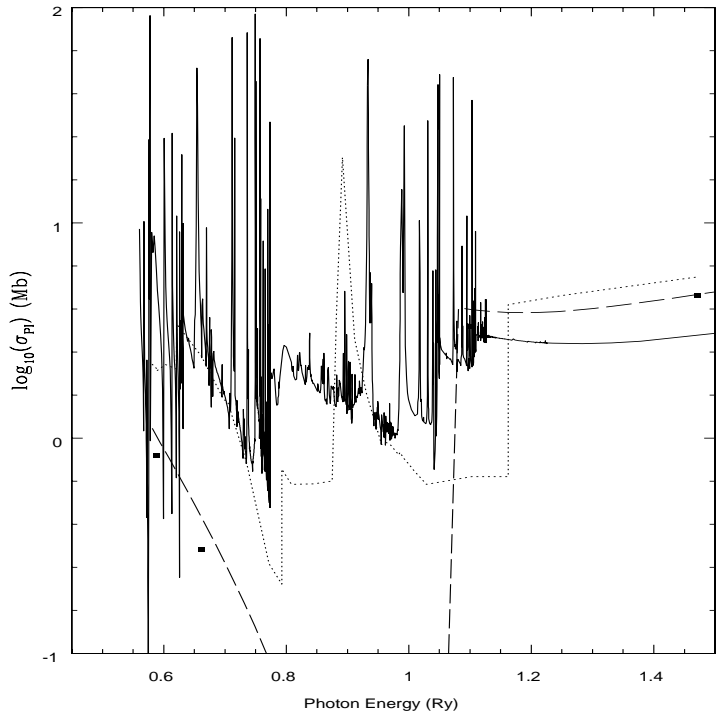,width=4in,height=4in}
\caption{\small Photoionization cross section of the ground state $3d^64s^2\
^5D$ of Fe~I. The full curve represents the R-matrix results of 
Bautista and Pradhan (1995); the dotted curve shows the results of 
Kelly and Ron (1972) and Kelly (1972) using the many-body perturbation
method; the dashed curve and the square dot represent the central-field
results of Verner \etal (1993) and Reilman and Manson (1979)
respectively.}
\end{figure}

Central field type calculations of photoionization cross sections have
been
carried out by Reilman and Manson
(1979), who adopted Hartree-Slater wave functions (Herman and Skillman
1963), and Verner \etal (1993), using the Dirac-Slater potential
(Slater 1960; Band \etal 1979).
These cross sections are reasonable accurate at high photon energies,
but
often give poor results near the threshold, particularly where CI is
strong. For example, in the photoionization of Fe~I the central field
cross section from the threshold to the $3d$ subshell is underestimated
by more than three orders of magnitude, as seen in Figure 2 (Bautista
and Pradhan
1995).
Similar errors of up to orders of magnitude
are found in Fe~II through Fe~V (Bautista and Pradhan 1998).

Other problems with central field cross sections are that they neglect 
resonances and, because only degenerate subshells of equivalent 
electrons rather than individual energy levels are considered, 
the utility of these cross sections in detailed spectral modeling is
limited.

\subsection{The Gaunt factor}            

Burgess, Seaton, and Van Regemorter (Burgess 1961; Seaton 1962, Van
Regemorter 1962) suggested an approximate formula to obtain near
threshold collision
strengths for optically allowed transitions. The formula is based on the
Kramers approximation and the Gaunt factor, $g$, and is usually accurate
within a factor of a few. 
Later, the formula was modified to replace $g$, which is a varying
function with energy,  by an empirical parameter $\bar g\sim 0.2$.  The
formula is
\begin{equation}
\Omega(i,i')={8\pi\over 3\sqrt(3)}S(i,i'){\bar g},
\end{equation}
where $S(i,i')$ is the line strength, which relates to the 
absorption oscillator strength as $g_{i'}f_{i'i}=(2/3)(E_{i'}-E_i) S(i,i')$

At high energies $\bar g$ increases logarithmically and the collision 
strength takes the form (Seaton 1962) 
\begin{equation}
\Omega(i.i')={4\over 3}S(i,i')\ln [4k^2/(r_0\Delta E)^2],
\end{equation}
where $\Delta E=E_{i'}-E_i$, and $r_0$ is the distance of closest approach
given by
\begin{equation}
{2z\over r_0}-{l(l+1)\over r_o^2}+k^2=0.
\end{equation}

\subsection{The Coulomb-Born approximation}

In the Coulomb-Born (CB) approximation $V_{ii'}$ is replaced by its
asymptotic form at large $r$, $2z/r$, where $z=Z-N+1$. In doing this the
solutions to the scattering problem are known analytically. The CB
approximation is best for highly charged ions and transitions not
affected by channel coupling and large
partial waves (e.g. Seaton 1975; Van Regemorter 1960).

\subsection{The Distorted Wave approximation}

The Distorted Wave (DW) method assumes the coupling between different 
target states to be weak ($V_{ii'}=0$, for $i\neq i'$ in
Equation 19). Then, the system of coupled equations is reduced to 
\begin{equation}
\{-(\nabla^2+k_i^2)+V_{ii}\} \theta_i=0.
\end{equation}
Some of the most refined forms of the method were developed by Saraph,
Seaton, and Shemming
 (1969) and Eissner and Seaton (1972) including exchange, which 
is important in many cases, like in intercombination transitions. 
In addition, {\it correlation functions}, $\chi_i$, which are
wavefunctions of the $(N+1)$-electron system, were introduced to
modify the wavefunctions as 
\begin{equation}
\Psi=\sum_i
\phi_i(x_1,...,x_N)\theta_i(x_{N+1})+\sum_{j=1}c_j\chi_j(x_1,...,x_{N+1}),
\end{equation}
where the coefficients $c_j$ are determined variationally.
This form of expansion of the wavefunction is known as the
close-coupling (CC) expansion (Seaton 1953).            

The DW method usually gives accurate collision strengths in the absence
of resonances for systems more than a few 
times ionized (see paper by Bhatia in this volume). Hershkowitz and Seaton (1973) 
showed that the DW method could also provide information about bound 
states of the $(N+1)$-electron system, and hence obtain resonance
structures. However, for detailed calculations of cross sections 
including resonances it is usually preferred to use the R-matrix method.

\subsection{The R-matrix method} 

The R-matrix method is the most sophisticated of the techniques 
discussed here and is also the most accurate. The method takes into   
account nearly all of the physical effects that contribute to cross sections 
for astrophysical applications and is applicable to all kinds of ions,
from neutral to highly ionized species. Of course, with
the increased 
complexity of the calculations the R-matrix method can be computationally
very intensive.

The R-matrix theory starts by dividing the configuration space by 
a sphere of radius $a$ centered on the target nucleus. In the internal
region, $r< a$, where $r$ is the relative coordinate of the free      
electron, electron exchange and correlation between the scattered 
electron and the target are important.
Thus a CC expansion
like that in Equation (27) is adopted for the system. In the external 
region, $r> a$, electron exchange between the free electron and 
the target can be neglected if $a$ is large enough to contain the 
charge distribution of the target. Then the scattered electron moves 
in the long-range multipole potentials of the target ion. The internal and 
external regions are linked by the R-matrix on the boundary, $r=a$.

Following the theory from Burke \etal (1971), and Burke and Robb (1975),
the total wavefunction $\Psi$ in the inner region for any energy $E$ 
can be written in terms of the basis states set $\{\psi_k\}$ as 
\begin{equation}
\Psi=\sum_k A_{Ek}\psi_k
\end{equation}
where the $\psi_k$ functions are energy independent and are expanded as shown in Equation (27), and the energy dependence is carried through the
$A_{Ek}$ coefficients.

Then, the $R$ matrix is defined as
\begin{equation}
R_{ij}(E)={1\over 2a}\sum_k {w_{ik}(a)w_{jk}(a)\over E_k-E},
\end{equation}
where
\begin{equation}
{1\over r}w_{ik}(r)=<\chi_i|\psi_k>
\end{equation}
This $R$ matrix is the basic solution of the electron-scattering problem
as it allows one to determine the atomic structure of the $(N+1)$
system, 
the collision strengths, and photoionization cross sections. 
The $R$ matrix, the amplitude of the wavefunctions at the $r=a$
boundary,
and the poles $E_k$ are obtained from the eigenvalues and eigenvectors
of the Hamiltonian matrix. 
Further, quantitative 
results for physical parameters of interest are obtained 
by matching the solutions with those in the outer 
region which are 
 known analytically.

One important point to note is that in the R-matrix method the
solution inner region is obtained only once, then 
cross sections for any number of energy points are readily available.
Furthermore, although the basic computations in this method are 
lengthly, it turns out to be a very efficient technique as it 
provides results for large numbers of points which allow 
complex resonance structures in the cross sections to be delineated.

The R-matrix method has been implemented in the RMATRX package of codes
(Berrington, Eissner, and Norrington 1995) 
which has been widely used by several groups like the Opacity Project,
that carried out extensive atomic structure and photoionization
calculations of astrophysically important ions,  and the Iron
Project, and the RmaX Project (see papers by Pradhan and Berrington in
this volume).

\subsection{Important physical effects}

There are several specific physical effects that should be taken into
account 
when calculating or evaluating cross sections for photoionization or
collisional excitation/ionization. These are: target CI, 
resonances, convergence of the wavefunction expansion, convergence of 
partial waves expansion, relativistic effects, and radiation damping.

\subsubsection{Configuration interaction}

The first requirement in any scattering calculation is a good
representation of the target, i.e. accurate wavefunctions for the 
target ion. Such representation usually requires the inclusion of CI 
in the atomic structure model, which as we saw in Section 2.5.1 affects
the calculated energy levels and oscillator strengths of the ion.
Furthermore, comparisons between calculated and experimental energies
and oscillator strengths of the target are important indicators of the
quality of the target representation and the overall accuracy of the 
obtained cross sections.

\subsubsection{Resonances}

Resonances are an important part of collisional excitation/ionization 
and photoionization cross sections. Physically, resonances occur when 
the incoming particle (electron or photon) with just the right kinetic
energy gets trapped into an autoionizing state\footnote{Autoionizing
states are compound states of the (electron+ion) system located      
above
the ionization potential. These states result from the excitation
of two or more
electrons of the system or by excitation of inner-shell
electrons.} 
 of the $(N+1)$-electron
system. Then, as the  electron remains trapped for a time before
autoionization occurs, the time delay  yields a phase
shift in the wavefunction that manifests itself in sharp peaks or throughs
in the cross sections (resonances).

\begin{figure}[Htb]
\hskip 1in\psfig{file=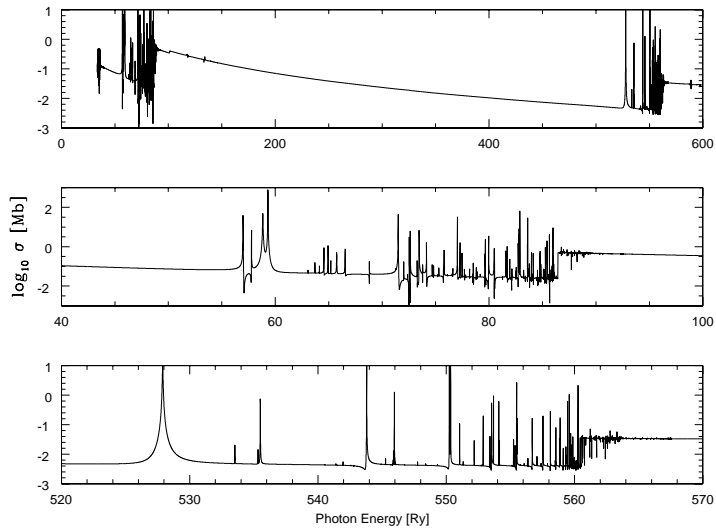,width=4in,angle=-90}
\caption{\small Photoionization cross section of the Fe~XV
$2p^63s^2(^1S)$ ground state from the R-matrix calculation of Bautista
(2000a). The top panel shows the full cross sections and the lower
panels
present close-ups of the cross section around the L- and K-inner shell
thresholds.}
\end{figure}

Resonances appear as Rydberg series converging onto the various
excitation thresholds of the target. 
In complex ions with many levels close in energy the series of 
resonances often overlap and interference effects can occur which 
makes the resonances inseparable from the background cross section.
An example of this is the photoionization cross section of the 
ground state of Fe~I (Figure 2). The R-matrix cross section 
(Bautista and Pradhan 1995) given by the solid line is over 
three orders of magnitude greater than the central field results
without resonances. Similar discrepancies with respect to central field
photoionization cross section are found for all iron ions up to Fe~V
(Bautista and Pradhan 1998). Like in the case of photoionization near the 
inner $3d$ subshell of Fe~I, 
resonances are a general phenomenon near inner-shell ionization
thresholds. 
Figure 3 shows the photoionization cross section
of Fe~XV from the first ionization threshold to just above the 
K-shell threshold (Bautista 2000a). Complex resonances structures 
are seen 
converging onto both L and K-shell thresholds. 
The anhancement of the cross section due to resonances near the L-shell
threshold was first pointed out by Hanque and Pradhan (1999) 
These resonance can change the appearance of the thresholds 
from sharp edges to more complex and continuous transitions. The 
resonances also enhance the photoionization rates. For example,
assuming a power law continuum of the kind $F\propto E^{-\alpha}$
for $E$ from 500 to 700 Ry, the resonances near the K-shell threshold 
of Fe~XV enhance the ionization rate of this ion 
by factors of 1.7 for $\alpha=1$, 1.8 for $\alpha=2$, and 2.9 
for $\alpha=3$. 

\begin{figure}[Htb]
\hskip 1in  \psfig{file=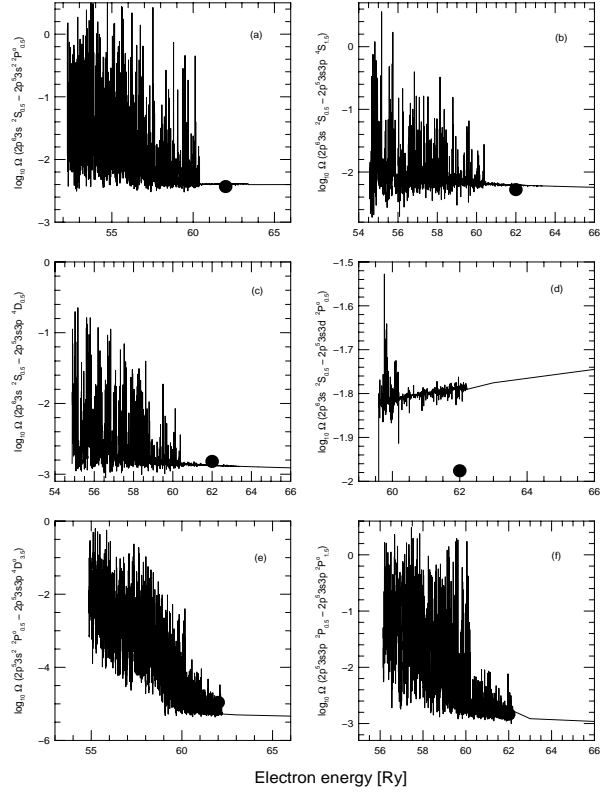,width=4in,height=5in}
  \caption{\small Comparison between R-matrix collision
strengths from Bautista (2000b) and distorted wave results by
Phillips \etal (1997).}
\end{figure}

Resonances in the collision strengths are known to enhance the 
excitation rates by up to several factors in the case of valence 
electron excitation.
In the case of inner-shell excitation, Bautista (2000b) carried out 
R-matrix calculations for Fe~XVI and compared the results with 
earlier DW results. In the region free of resonances the agreement
is generally very good, but huge numbers of near threshold narrow resonances 
are found from the R-matrix calculation (see Figure 4). These resonances
enhance the effective collision strengths by up to three order of
magnitude, as shown in Table 1.

\begin{table}[htb]           
\caption
{Comparison between effective collision strengths from 
Bautista (2000b) 
and the DW results of Phillips \etal
(1997).
The results are given for temperatures of $5.10^4$,
$10^5$ and $10^6$ K. The last column indicates the maximum factor of
enhancement over the DW results.}
\begin{center}\small      
\begin{tabular}{llccccr}
\tableline
Lower level& Upper Level & $\Upsilon(5.10^4$K)&$\Upsilon(
10^5$K)&$\Upsilon(10^6$K) & Phillips \etal& max. ratio \cr
\tableline
$2p^63s  ~^2S_{0.5}
$&$2p^53s^2~^2P_{1.5}^o$&4.02(-2)&4.37(-2)&1.99(-2)&3.70(-3
)&12\cr

&$2p^53s^2~^2P_{0.5}^o$&2.01(-2)&2.18(-2)&9.96(-3)&1.96(-3
)&11\cr
                      &$2p^53s3p~^4D_{2.5}
$&4.10(-2)&3.43(-2)&1.24(-2)&2.10(-4
)&236\cr
                      &$2p^53s3p~^4D_{3.5}
$&2.92(-2)&2.36(-2)&5.36(-3)&1.10(-5
)&2660\cr
                      &$2p^53s3p~^4P_{2.5}
$&2.17(-2)&1.87(-2)&6.76(-3)&6.60(-5
)&380\cr
                      &$2p^53s3p~^2D_{2.5}
$&1.18(-1)&9.92(-2)&2.74(-2)&1.21(-3
)&105\cr
$2p^53s3p~^4S_{1.5}  $&$2p^53s3p~^4P_{1.5}
$&3.05(-1)&3.46(-1)&1.22(-1)&5.38(-3
)&64\cr
$2p^53s3p~^4D_{2.5}  $&$2p^53s3p~^2D_{1.5}
$&5.93(-2)&5.66(-2)&1.79(-2)&2.03(-4
)&292\cr
$2p^53s3p~^4D_{0.5}  $&$2p^53s3p~^4D_{3.5}
$&5.41(-1)&4.61(-1)&1.24(-1)&3.40(-3
)&159\cr
$2p^53s3p~^4D_{3.5}  $&$2p^53s3p~^2P_{0.5}
$&1.69(-2)&1.52(-2)&6.83(-3)&6.12(-4
)&30\cr
$2p^53s3p~^4D_{1.5}  $&$2p^53s3p~^2D_{1.5}
$&5.23(-2)&4.95(-2)&1.57(-2)&3.24(-4
)&161\cr
$2p^53s3p~^4P_{2.5}  $&$2p^53s3p~^2P_{1.5}
$&1.05(-1)&9.25(-2)&3.84(-2)&3.22(-3
)&38\cr
                      &$2p^53s3p~^2D_{1.5}
$&4.84(-2)&4.15(-2)&1.31(-2)&1.47(-4
)&329\cr
$2p^53s3p~^4P_{1.5}  $&$2p^53s3p~^2D_{1.5}
$&4.02(-2)&3.47(-2)&1.18(-2)&2.82(-4
)&143\cr
$2p^53s3p~^4P_{0.5}
$&$2p^53s3d~^4F_{4.5}^o$&1.04(-2)&9.27(-3)&3.03(-3)&7.00(-5
)&162\cr
                      &$2p^53s3p~^2D_{2.5}
$&1.47(-1)&1.14(-1)&3.11(-2)&9.72(-4
)&224\cr
                      &$2p^53s3p~^2P_{0.5}
$&2.19(-1)&1.93(-1)&5.23(-2)&1.43(-3
)&154\cr
$2p^53s3p~^2P_{1.5}  $&$2p^53s3p~^2D_{2.5}
$&3.46(-1)&2.80(-1)&7.80(-2)&2.98(-3
)&161\cr
                      &$2p^53s3p~^2P_{1.5}
$&6.01(-1)&5.40(-1)&1.51(-1)&1.85(-3) & 325\\
\tableline
\end{tabular}
\end{center}
\end{table}

\subsubsection{Convergence of the partial waves expansion}

It is an standard approach to expand the collision strengths in partial
waves from 0 to infinity from every possible value of the angular momentum of the 
free electron. Clearly, infinite expansion terms cannot be computed
in practice and one takes only the lowest dominant partial waves 
and extrapolate through infinity. The convergence of the partial
waves expansion is usually rapid for forbidden transitions, but 
it can be quite slow for allowed transitions. Also, the number of 
partial waves needed for convergence increases with increasing 
energy of the free electron. 
Thus, the convergence of the partial waves expansion becomes a difficult
practical problem, and a possible source of error in the collision
strengths for highly ionized systems for which  very high 
collision energies need to be considered (e.g. Burke and Seaton 1986; Chidichimo 1988, 1989; Eissner \etal 1999)

\subsubsection{Convergence of the close coupling expansion}

The general form of the CC expansion for the radial
wavefunction $\Psi$ in terms of a $N$-electron target basis $\chi_i$
and the scattering electron function $\theta_i$ is
\begin{equation}
\Psi=\sum_i^\infty \chi_i\theta_i+\int_\epsilon \chi_\epsilon
\theta_\epsilon.
\end{equation}
In the DW and the standard R-matrix approach, however, the integral 
term that accounts for the target continuum is neglected
or replaced by a discrete sum over bound correlation functions 
(Equation 27) 
and the sum over the infinite number of target states is truncated 
to a small number of strongly coupled states. This approximation 
is usually good when considering valence electron excitations among 
the lowest energy levels of the ion. However, for excitations to 
highly excited levels and inner-shell excitations the convergence 
of the CC expansion must be looked at in great detail.
For example, Sawey and Berrington (1993) showed that in order to obtain 
accurate collision strengths for the $n=4$ levels of He~I one must 
include target states in the CC expansion up to at least $n=5$.
Otherwise, if the CC expansion is too small the obtained collision 
strengths tend to be severely overestimated. 

The use of pseudo-orbitals to accelerate the convergence of CC 
expansions was first introduced by Burke, Gallagher, and Geltman (1969).
Recant developments are 
the convergent close-coupling (CCC) and R-matrix 
with pseudo-states (RMPS) methods (see the review by Gorczyca 
\etal\  in this volume).

\subsubsection{Relativistic effects}

If relativistic corrections are small, one can carry out the scattering
calculation in LS coupling and then perform an algebraic
transformation of the $K$ (reactance) matrices to fine structure 
(Saraph 1972, Luo and Pradhan 1990). 

As the relativistic effects become significant one must allow for the 
fine structure splitting of the target by using the so-called 
term-coupling-coefficients (TCCs) to diagonalize the Hamiltonian with
the inclusion of relativistic corrections.

For more complete and proper treatament of relativistic effects      
it is necessary to use either the Breit-Pauli or the Dirac formulations.
The Breit-Pauli operators have been implemented into the RMATRX 
package of codes by the Iron Project (Hummer \etal 1993). One limitation 
with relativistic calculations is that by splitting the structure of 
the core into fine structure the size of the computation 
often exceeds the capacity of modern vectorized supercomputers. 
Perhaps, this problem may be solved in the near future with the use 
of massively parallel computers. 

\subsubsection{Radiation damping}

Radiation damping of resonances can be an important effect for highly 
charged 
ions. When the radiative de-excitation rates of autoionizing states 
are comparable to the autoionization rates the resonances 
associated with these states become damped. 
In the case of photoionization damping of resonances occurs because 
as the target is photo-excited to an autoionizing state 
radiative de-excitation competes with autoionization.
In collisional excitation damping of resonances indicates a 
competition between dielectronic recombination and autoionization.

Radiation damping in electron-ion scattering were studied by Pradhan 
(1981, 1983a, 1983b) using 
the branching ratios between autoionization and radiative rates
and more recently by Zhang and Pradhan (1995) employing the detailed Bell and 
Seaton (1985) theory of dielectronic recombination. 
Radiation damping was found to be important in reducing the strength
of resonances in boron-like Fe~XXII and
helium-like Fe~XXV. As expected, radiation damping increases with the 
principal quantum number $n$ within a give Rydberg 
series of resonances. This is because autoionization rates typically
decrease as $n^{-3}$ while radiative rates remain approximately constant.

Robicheaux \etal (1995) and
Gorczyca and Badnell (1996) have studied radiation damping in 
photoionization cross sections. 
However, in practical applications of photoionization and
and scattering the overall radiation damping
effect may be not so great in most cases. This is because while the 
higher $n$ resonances can be highly damped, the lower resonances in the
series, which are usually dominant, are not affected significantly, 
except in highly charged ions (Pradhan and
Zhang 1997; Robicheaux 1998).

\section{Conclusions}

A number of methods for atomic structure and electron-ion scattering
calculations have been developed, which are able to provide atomic 
data for spectroscopic applications. These methods vary in complexity and 
accuracy of the results in such a way that one must often
decide between promptly available and highly accurate data to fulfill 
current spectroscopic needs.      
In the case of X-ray spectroscopy most of the atomic data currently 
available have been produced using some of the simplest techniques 
described here, while improved data  such as that from the IRON Project 
(see review by Pradhan) is slowly been calculated. 
In some cases, new more elaborate calculations will help in refining 
the spectral models, while in other cases the new atomic data could 
vastly change the previous results leading to dramatic changes 
in the spectral models. 
Therefore, it is of prime importance to researchers trying to 
analyze and/or model spectra to have at least some basic understanding
of the quality of the data bing used. 


\def\r{\leftskip20pt \parindent-20pt \parskip0pt}
\def\apj{{Astrophys. J.}\ }
\def\apjs{{Astrophys. J. Supp. Ser.}\ }
\def\aj{{Astron. J.}\ }
\def\adndt{{Atom. Data and Nucl. Data Tables}\ }
\def\pasp{{Pub. Astron. Soc. Pacific}\ }
\def\mn{{Mon. Not. R. Astr. Soc.}\ }
\def\aap{{Astron. \& Astrophys.}\ }
\def\aasup{{Astron. \& Astrophys. Suppl.}\ }
\def\baas{{Bull. Amer. Astron. Soc.}\ }
\def\cpc{{Comput. Phys. Commun.}\ }
\def\jqsrt{{J. Quant. Spectrosc. Radiat. Transfer}\ }
\def\jpb{{J. Phys. B: Atom., Mol. \& Optic. Physics}\ }
\def\pra{{Phys. Rev. A}\ }
\def\adndt{{At. Data. Nucl. Data Tables}\ }

\end{document}